\documentclass[aps,prl,floatfix,twocolumn]{revtex4}
\usepackage{graphicx,amsmath,amssymb,subfigure,color,verbatim,array}
\usepackage{yllmath}
\newcommand{\pppi}{\boldsymbol{\pi}}

\begin{document}
\title{Large Range of Stability of Larkin-Ovchinnikov States for Imbalanced Fermi Gases in Optical Lattices}
\author{Y.~L.~Loh}
\author{N.~Trivedi}
\affiliation{Department of Physics, The Ohio State University, 191 W Woodruff Avenue, Columbus, OH  43210}
\date{\today}
%\date{\red{Started 2008-11-30, touched 2009-6-1, compiled \today}}
\begin{abstract}
We show that Larkin-Ovchinnikov (LO) states with modulated superfluid order parameters have a considerably larger range of stability in a lattice than in the continuum.  We obtain the phase diagram for the 3D cubic attractive Hubbard model with an unequal population of up and down fermions using the Bogoliubov-de Gennes fully self-consistent method.  
We find a strong modulation of the local polarization that should provide a distinct signature for detection of the LO phase.  The shell structure in the presence of a trap generates singularities in the density at the phase boundaries and provide additional evidence for the LO phase.
Depending on specific parameters,  the LO ground state occurs over a large range of population imbalance, involving 80$\percent$ of the atoms in the trap, and can exist up to an entropy $s\sim 0.5 k_B$ per fermion.
\end{abstract}
\maketitle
	
%\tableofcontents
%@@@@@@@@@@@@@@@@@@@@@@@@@@@@@@@@@@@@@@@@@@@@@@@@@@

%@@@@@@@@@@@@@@@@@@@@@@@@@@@@@@@@@@@@@@@@@@@@@@@@@@
An imbalanced population of fermions with two hyperfine states and interacting via attractive interactions offers the exciting possibility of observing superfluidity with a spatially modulated order parameter. For small imbalance the ground state is a BCS/BEC superfluid state with paired fermions, but for large imbalance the ground state is a polarized Fermi liquid.\cite{chandrasekhar1962,clogston1962}
At intermediate polarizations, mean-field theories predict Fulde-Ferrell-Larkin-Ovchinnikov (FFLO) states with a spatially modulated superfluid order parameter
\cite{fulde1964,larkin1964} that is a compromise between pairing and polarization.
There is controversy over whether the FFLO state exists.  1D systems only allow a quasi-long-range-ordered version of FFLO, whereas in 2D and 3D continua, FFLO only occupies a sliver of the phase diagram and is vulnerable to fluctuations.  So far, ordered FFLO has not been observed except in some reports on layered organic and heavy-fermion superconductors\cite{radovan2003}. 
Both BCS and polarized states have been observed in imbalanced cold fermionic gases 
% in traps 
%in the continuum~
\cite{partridge2006,shin2006}, but the LO phase has so far remained elusive.

% OTHER THEORY PAPERS ON LO:  \cite{casula:033607}
% I'm not citing

\begin{figure}
\includegraphics[width=8cm]{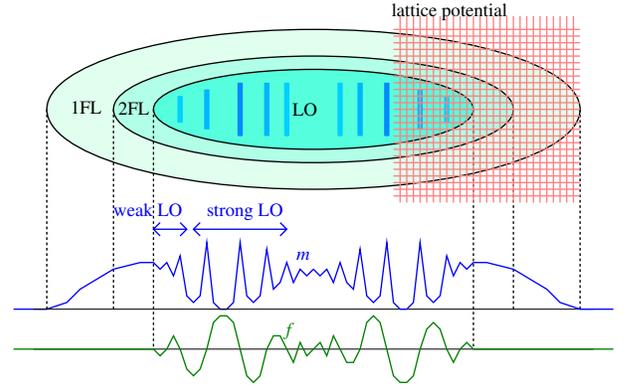}
\caption{
Schematic depiction of an imbalanced fermion gas in an optical lattice in a cigar-shaped trap, corresponding to slice (i) in Fig.~\ref{U-6}.  There are regions of strong LO order consisting of domain walls (DWs), in which the excess majority-spins fermions are confined.  The polarization in each DW can be about $0.25/a^2$ fermions per unit area, the DW thickness about $3a$, and the DW spacing about $6a$ (where $a$ is the lattice constant).
}
\label{schematic}
\end{figure}

In this paper we 
study the full phase diagram of the cubic lattice Hubbard model.
We use approaches based on variational mean-field theory in six channels, which
includes Bogoliubov-de Gennes (BdG) and Hartree corrections. Our main results are:
(1) The LO phase occupies a large region of the phase diagram between the BCS and polarized Fermi liquid (2FL) phases.
%(2) A defining characteristic of the LO phase it that the order parameter changes sign evolving from a domain wall structure at low fields to a sinusoidal variation at higher fields. This is accompanied by a
%variation of the polarization which also evolves from being concentrated at the domain walls to becoming sinusoidal.
(2) With increasing field (or imbalance), the BCS state becomes unstable to an LO phase consisting of domain walls at which the order parameter changes sign.  The polarization is confined to these domain walls.  At higher fields the domain wall structure evolves into a sinusoidal variation of the order parameter accompanied by a polarization variation at twice the wavevector.   We suggest that the most promising way to detect the LO phase is to focus on this spatial variation of the polarization.
(3) The momentum distribution functions $n_\sigma(\kkk)$ in the LO phase shows distinct features that break the lattice symmetry, such as Fermi arcs, Fermi pockets, and blocking regions, unlike in the BCS/BEC state.
(4) Depending on parameters, the LO phase can exist below an entropy $s\sim 0.5k_B$.
(5) In an optical lattice in a shallow trap with appropriate parameters, LDA predicts that most ($>80\percent$) of the atoms participate in the LO phase!  See Fig.~\ref{schematic}.

%%%%%%%%%% FIGURE: SCHEMATIC %%%%%%%%%%%%%%%%%
%%%%%%%%%% END FIGURE %%%%%%%%%%%%%%%%%

%==============================================================================

%==============================================================================
\paragraph{Model and methods:}
The Hubbard Hamiltonian is given by
  \begin{align}
  H &=
  \sum_{\kkk\sigma} \xi_{\kkk} (n_{\kkk\sigma} - \half)
	- \sum_{\rrr\sigma} \mu_\sigma (n_{\rrr\sigma} - \half)
	\nonumber\\&{}~~~
  + U \sum_{\rrr} (n_{\rrr\up} - \half)(n_{\rrr\dn} - \half)
  \end{align}
where
$  \xi_{\kkk} = -2(\cos k_x + \cos k_y + \cos k_z)
$
is the dispersion relation on the cubic lattice for nearest-neigbor hopping, 
$\sigma=\pm 1$ labels (hyperfine) spin states,
$ n_{\rrr\sigma} = \cdag_{\rrr\sigma} \cccc_{\rrr\sigma}
$
are number operators,
$  \mu_{\sigma} = \mu + \sigma h
$
are the chemical potentials for the two spin species, and $U$ is the local pairwise Hubbard interaction.  The hopping $t$ is the unit of energy.  We use the convention that repulsive $U$ is positive.
We find it convenient to work in terms of the average chemical potential $\mu$ and the Zeeman field $h=2(\mu_\up-\mu_\dn)$.  The observables of interest are the density 
$n_\rrr = \half(n_{\rrr\up}+n_{\rrr\dn})$, 
imbalance
$m_\rrr = \half(n_{\rrr\up}-n_{\rrr\dn})$, 
and pairing density %(anomalous equal-time Green function) 
$F_\rrr = \mean{c_{\rrr\dn} c_{\rrr\up}} = \frac{\Delta_\rrr}{-U}$.
Since $H$ has been written in terms of symmetrized densities $n_\sigma-\half$ measured with respect to half-filling, the thermodynamic quantities and phase diagram are invariant under sign changes of $\mu$ and $h$.
Furthermore, the ``Lieb-Mattis'' transformation (LMT) relates the repulsive and Hubbard model as described in Table \ref{t:LiebMattis}.

%%%%%%%%%% TABLE: LIEB-MATTIS MAPPING OF HUBBARD MODEL %%%%%%%%%%%%%%%%%
\begin{table}[!tb]
\begin{ruledtabular}
\begin{tabular}{cc}
$+U$ & $-U$ \\ 
$\mu$ & $h$ \\
$h$ & $\mu$ \\
\hline
$m_X(\qqq)$ & $\Re F(\qqq+\pppi)$ \\ 
$m_Y(\qqq)$ & $\Im F(\qqq+\pppi)$ \\ 
$m_Z(\qqq)$ & $n(\qqq)$ \\
Perpend. spin suscep. $\chi_X(\qqq)$ & Pairing suscep. $\Pi(\qqq+\pppi)$ \\
Parallel spin suscep. $\chi_Z(\qqq)$ & Compressibility $\kappa(\qqq)$ \\
\hline
$z$-$\pppi$-SDW  & $\pppi$-CDW \\
$x$-$\pppi$-SDW  & BCS/BEC superfluid \\
$z$-$\qqq$-SDW   & $\qqq$-CDW \\
$x$-$\qqq$-SDW   & $\qqq$-LO \\
$d$-wave superfluid  & exotic $d$-wave bond magnetism \\
\end{tabular}
\end{ruledtabular}
\caption{\label{t:LiebMattis}
Effect of the ``Lieb-Mattis transformation'' (LMT) on the parameters, observables, and phases of the Hubbard model on a bipartite lattice.  The LMT is a particle-hole transformation on the down spins with a $\pi$ phase shift for the B sublattice: $c_{A\dn} \rightarrow \cdag_{A\dn}$, $c_{B\dn} \rightarrow -\cdag_{B\dn}$.  It relates magnetic phases of the repulsive model to paired/density-ordered phases of the attractive model.  A $z$-$\pppi$-SDW has a spin density modulation in the $z$ direction at the antiferromagnetic wavevector $(\pi,\pi,\pi)$, whereas an $x$-$\qqq$-SDW has moments in the $x$ direction at some general wavevector $\qqq$; the latter maps to a LO (Larkin-Ovchinnikov) state, as illustrated in Fig.~\ref{real-space}.
}
\end{table}
%YL comments:
%Pierna theorem says Hubbard doesn't have FM, pi-lo, Charge Sep.  So, omit from table.
%%%%%%%%%%%%%%%%%%%%%%%%%%%%%%%%%%%%%%%%%%%%%%%%%%%%%%%%%%%%%%%%%%%%%%%

%==============================================================================
%\myhrulefill\section{Methods}
%==============================================================================
Our calculations are based on $\Tr \rho \ln \rho$ variational mean-field theory.  The Hubbard $U$ is approximated by $6N$ potentials, where $N$ is the number of sites: the local chemical potentials $\mu^\text{int}_\rrr$, Zeeman fields $\hhh^\text{tot}_\rrr$, and complex pairing potentials $\Delta_\rrr$.  This is equivalent to full BdG with Hartree corrections.  The trial Hamiltonian involves a $4N\times 4N$ matrix $\HHH$:
	\begin{align}
	&\hat H^\text{trial} = \hat H^\text{kin}	-	\half\times
\nonumber\\{}
	&
	\psmat{
		\cdag_{\rrr\up} \\
		\cdag_{\rrr\dn} \\ 
		\cccc_{\rrr\up} \\
		\cccc_{\rrr\dn} \\
	}
	\psmat{
		\mu+h_Z & -h_X+ih_Y & 0 & \Delta_R+i\Delta_I \\
		-h_X-ih_Y & \mu-h_Z & -\Delta_R-i\Delta_I & 0 \\
		0	& -\Delta_R+i\Delta_I  & -\mu-h_Z & h_X-ih_Y \\
		\Delta_R-i\Delta_I & 0 & h_X+ih_Y & -\mu+h_Z \\
	}^\text{tot}_\rrr
	\psmat{
		\cccc_{\rrr\up} \\
		\cccc_{\rrr\dn} \\ 
		\cdag_{\rrr\up} \\
		\cdag_{\rrr\dn} \\
	}
	.
	\end{align}
This trial Hamiltonian corresponds to a trial density matrix $\rho$ and thus to a variational free energy ``$\Tr \rho \ln \rho$'' %[cite Chaikin book],
	\begin{align}
								&
	\Omega_\text{var}
	= -\tfrac{T}{2} \tr \ln \left( 2 \cosh \tfrac{ \mathbf{H} }{2T} \right)
		+ \sum_\rrr \Big[
			U\Big( 
					\left|F_\rrr\right|^2 + {n_\rrr}^2 - \left|\mathbf{m}_\rrr\right|^2
			\Big)
						\nonumber\\&{}
		+ 2 \Big(
				\Re\Delta^\text{int}_\rrr \Re F_\rrr 
			+	\Im\Delta^\text{int}_\rrr \Im F_\rrr 
			+ \mu^\text{int}_\rrr n_\rrr 
			+ \hhh^\text{int}_\rrr \cdot \mathbf{m}_\rrr
		\Big) 
	\Big]
	\end{align}
where the ``pairing density' $F_\rrr=\mean{c_{\rrr\up} c_{\rrr\dn}}$, charge density $n_\rrr$, and spin density $\mathbf{m}_\rrr$ are elements of the matrix Green function
$	\GGG = -\half\tanh \tfrac{ \mathbf{H} }{2T}
$.
Differentiating $\Omega_\text{var}$ with respect to the variational parameters leads to the BdG and Hartree self-consistency conditions, and computing $\Omega_\text{var}$ itself is necessary to distinguish between various possible ground states.  
We use simplified methods (e.g., instability analysis) where applicable.

%==============================================================================
\paragraph{Phase diagrams:}
%==============================================================================

%%%%%%%%%% FIGURE: U=0 PHASE DIAGRAM, U=6 PHASE DIAG, TRAP PROFILES %%%%%%%%%%%%%%%%%
\begin{figure}
\subfigure[
$U=0$
]{
\includegraphics[width=3.5cm]{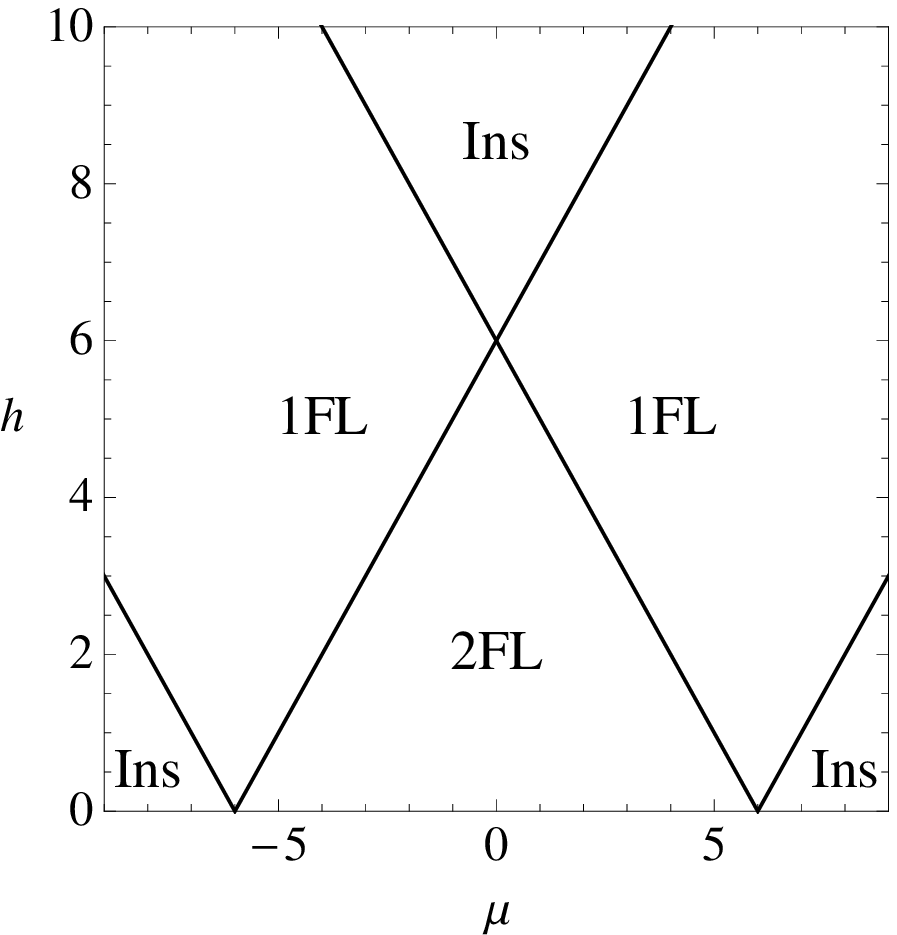}
\label{U0}
}
\subfigure[
$U=-6t$
]
{
\label{U-6}    %LABEL CAN'T BE REFERENCED FOR SOME SILLY REASON
\includegraphics[width=4.5cm]{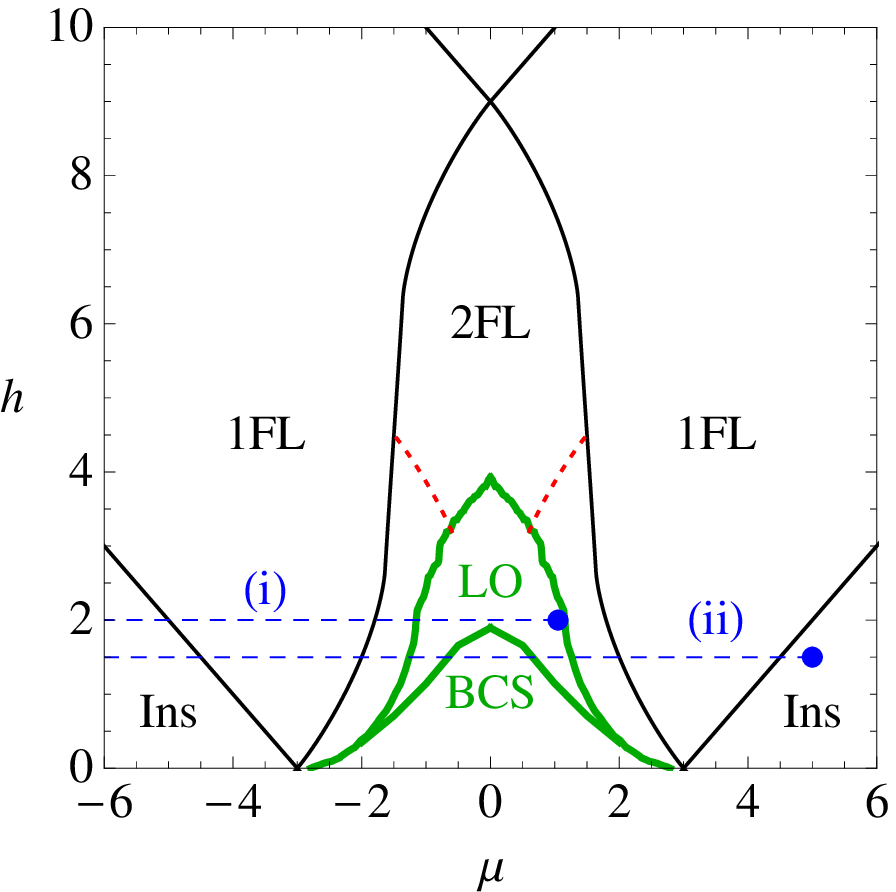}
}
\caption{
Phase diagrams of the cubic lattice Hubbard model %for $U=0$ in (a) and $U=-6t$ in (b) 
as functions of average chemical potential $\mu$ and Zeeman field $h$.
At large negative (positive) $\mu$ the system is an insulator with every site empty (doubly occupied).  At large $h$ it is a fully-polarized insulator.  
1FL represents a fully polarized Fermi liquid (half-metal).  2FL represents a two-component Fermi liquid.
BCS and LO represent superfluid states.  The black curves indicate Lifshitz transition boundaries corresponding to changes in Fermi surface topology.
The red curves are slivers of a CDW phase that only occur at extremely low temperature, and are irrelevant to experiments.
The dashed blue lines represent slices through the phase diagram corresponding to two different combinations of overall polarization and trapping potential; each blue dot indicates the chemical potential at the center of the trap.  Slice (i) corresponds to a large LO core surrounded by 2FL and 1FL shells, as illustrated in Fig.~\ref{schematic}.  Slice (ii) corresponds to an insulating core with two fermions per site, surrounded by seven shells.
}
\label{phase-diagram}
\end{figure}
%%%%%%%%%% END FIGURE %%%%%%%%%%%%%%%%%

The ground state phase diagram has three parameters: $\mu$, $h$, and $U$.  It is symmetrical under sign changes of $h$ and $\mu$.
Fig.~\ref{U0} shows the non-interacting ($U=0$) phase diagram.  
At finite attractive $\left|U\right|$, two superfluid phases (BCS and LO) appear in the phase diagram.  The size of the LO region grows with $\left|U\right|$ until $\left|U\right| \approx 5t$.  At large $\left|U\right|$, mean-field theory is affected by fluctuations.  Therefore we will focus on $U=-6t$ (see Fig.~\ref{U-6}).  
Note that $U_u = -7.91355t$ is the coupling where a bound state first appears in the lattice (the analogue of unitarity in the continuum), while the bandwidth is $12t$.
Hartree corrections cause the phase diagram to be squashed in the $\mu$-direction and elongated in the $h$-direction.
The main encouraging observation is that between the BCS state at $h=0$ and the polarized FL state at moderate $h$, there is a sizeable region where the ground state is an LO state.

%%%%%%%%%% FIGURE: REAL SPACE DENSITIES %%%%%%%%%%%%%%%%%
\begin{figure}[!h]
\centering
\subfigure[
	Weak LO
	%$8\times 10\times 10, U=-6, \mu=-0.8, h=2.0$
]{
	\includegraphics[width=0.45\columnwidth]{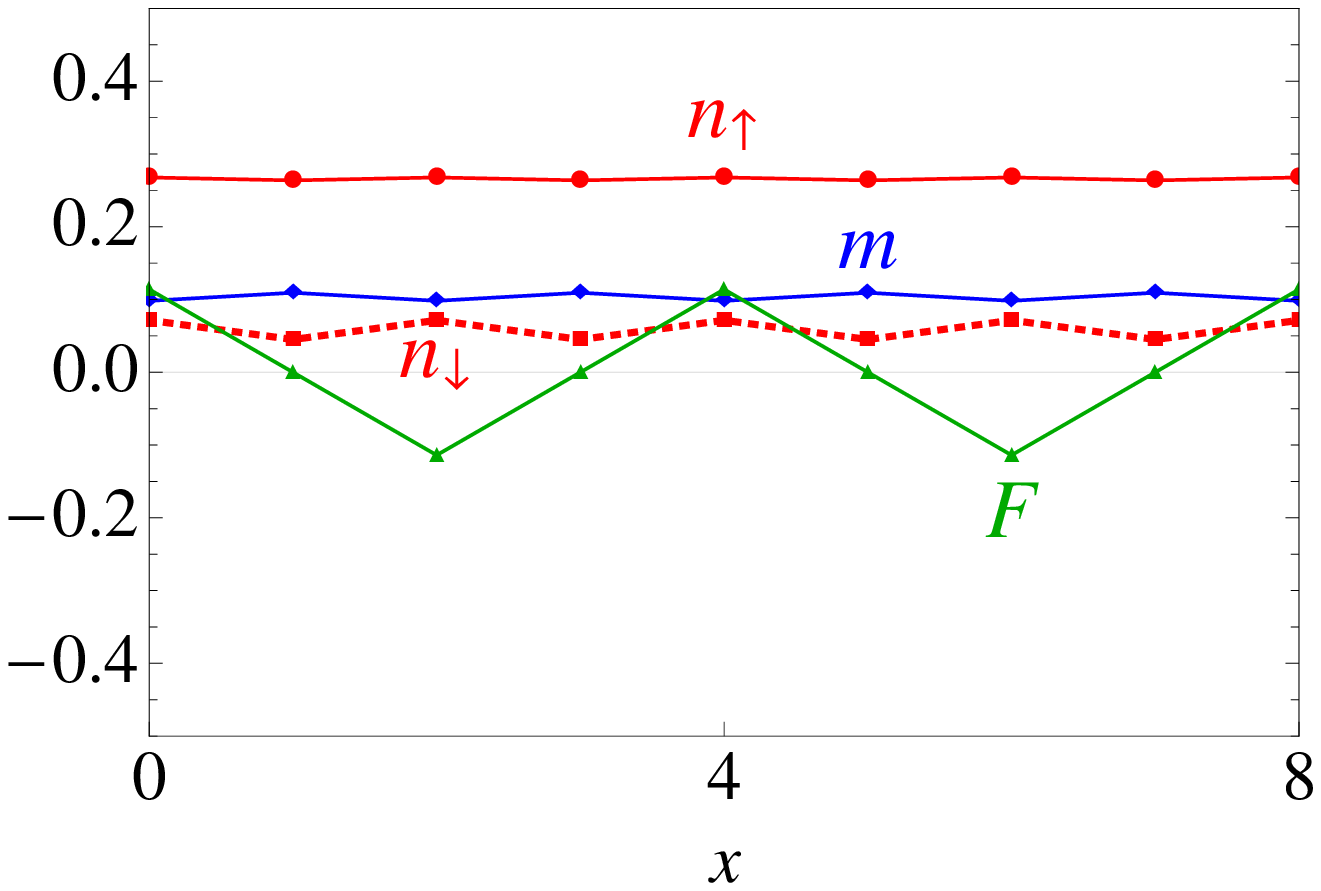}
}
\subfigure[
	Strong LO
	%$16\times 8\times 8, U=-6, \mu=-0.5, h=1.7$.  just above $h_{DW}=1.66$.
]{
	\includegraphics[width=0.45\columnwidth]{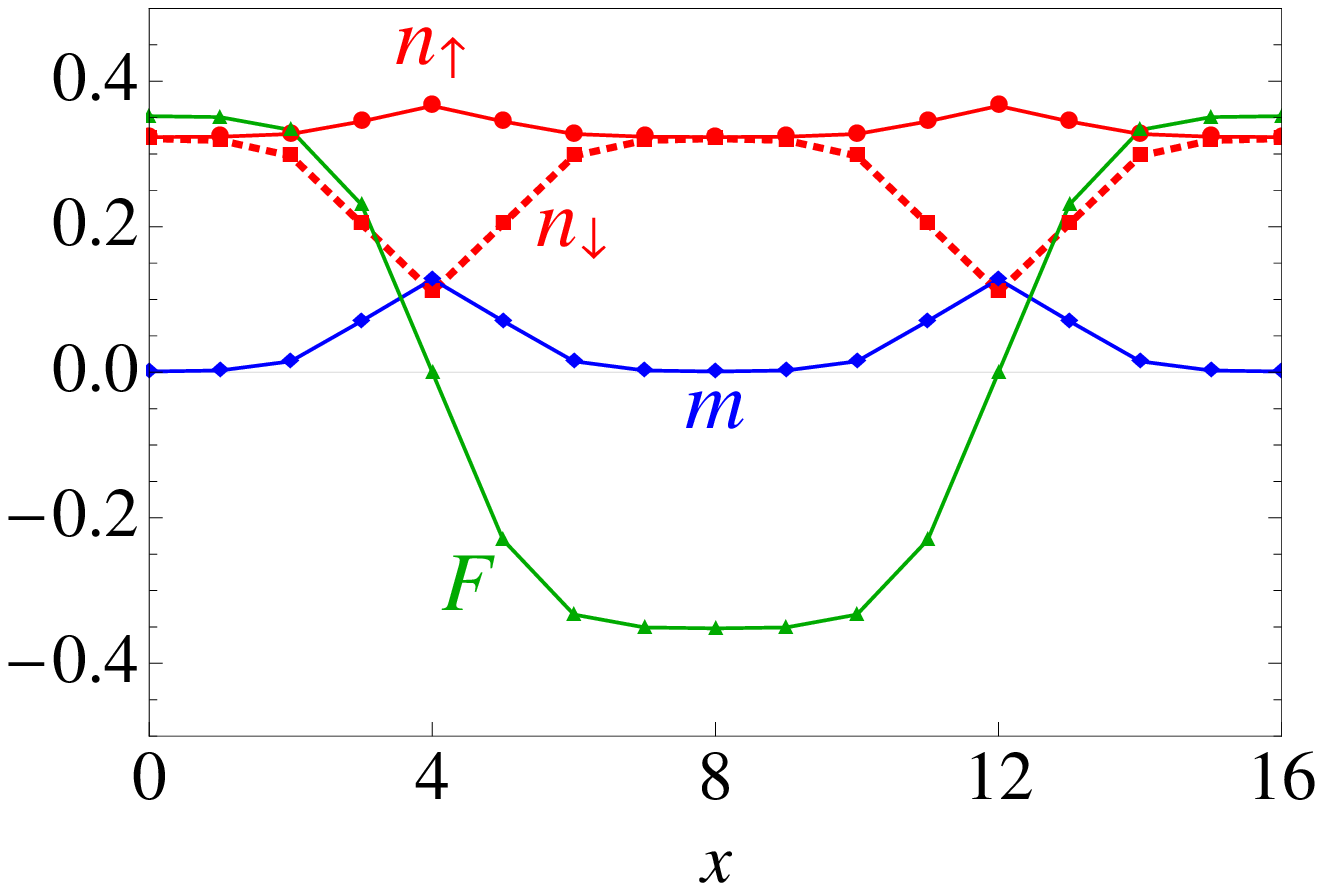}
	\label{strong-LO-real-space.eps}
}
\subfigure[
	Weak texture
	%$8\times 10\times 10, U=+6, h=-0.8, \mu=2.0$
]{
	\includegraphics[width=0.45\columnwidth]{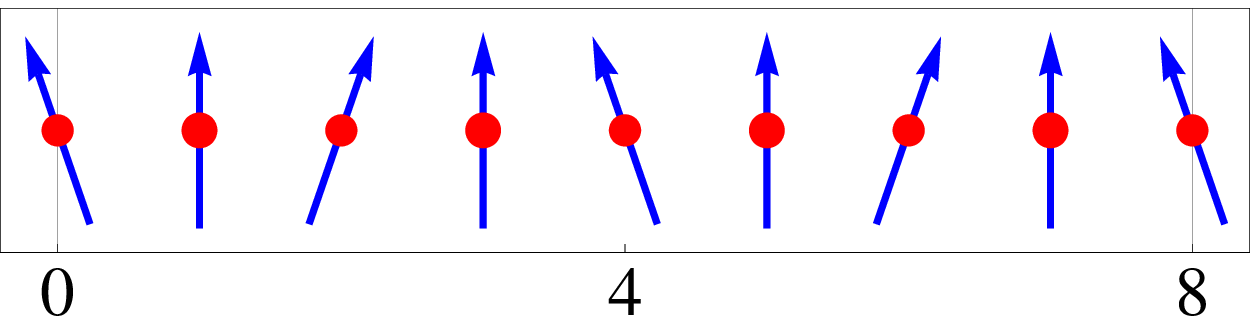}
}
\subfigure[
	Strong texture
	%$16\times 8\times 8, U=+6, h=-0.5, \mu=1.7$
]{
	\includegraphics[width=0.45\columnwidth]{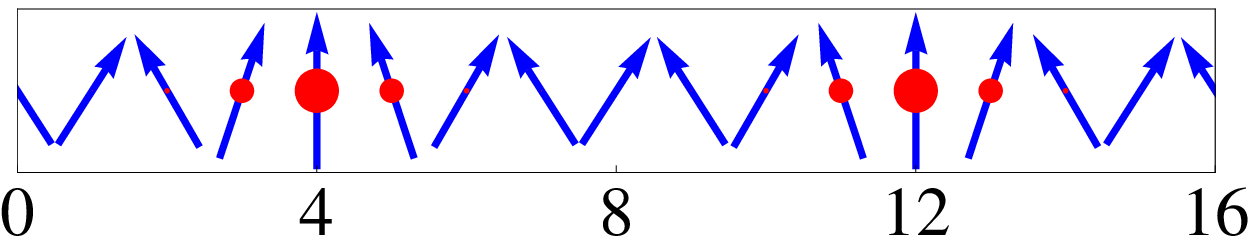}
}
\caption{
(a) Weak LO state with sinusoidal pairing density $F_\rrr \sim \cos\qqq\cdot\rrr$ accompanied by CDW and SDW at wavevector $2\qqq$.
(b) Strong LO state just above the critical field for domain wall penetration.  At each domain wall, the order parameter changes sign, and the polarization is finite due to occupation of Andreev bound states.
(c) Weakly textured state, in which the $z$-components of moments are nearly uniform and the $x$ components oscillate at wavevector $(\pi,\pi,\pi)-\qqq$.
(d) Strongly textured state (not to scale).  The red circles represent the hole density with respect to half filling, which is concentrated at the domain walls of the antiferromagnet.
Panels (a) and (c) are related to (b) and (d) by the LMT.
\label{real-space}
}
\end{figure}
%%%%%%%%%% END FIGURE %%%%%%%%%%%%%%%%%

%%%%%%%%%% FIGURE: MOMENTUM DISTRIBUTIONS %%%%%%%%%%%%%%%%%
\begin{figure}
\centering
\subfigure[$n_\up(k_x,k_y,k_z=0)$]{
\includegraphics[width=0.43\columnwidth]{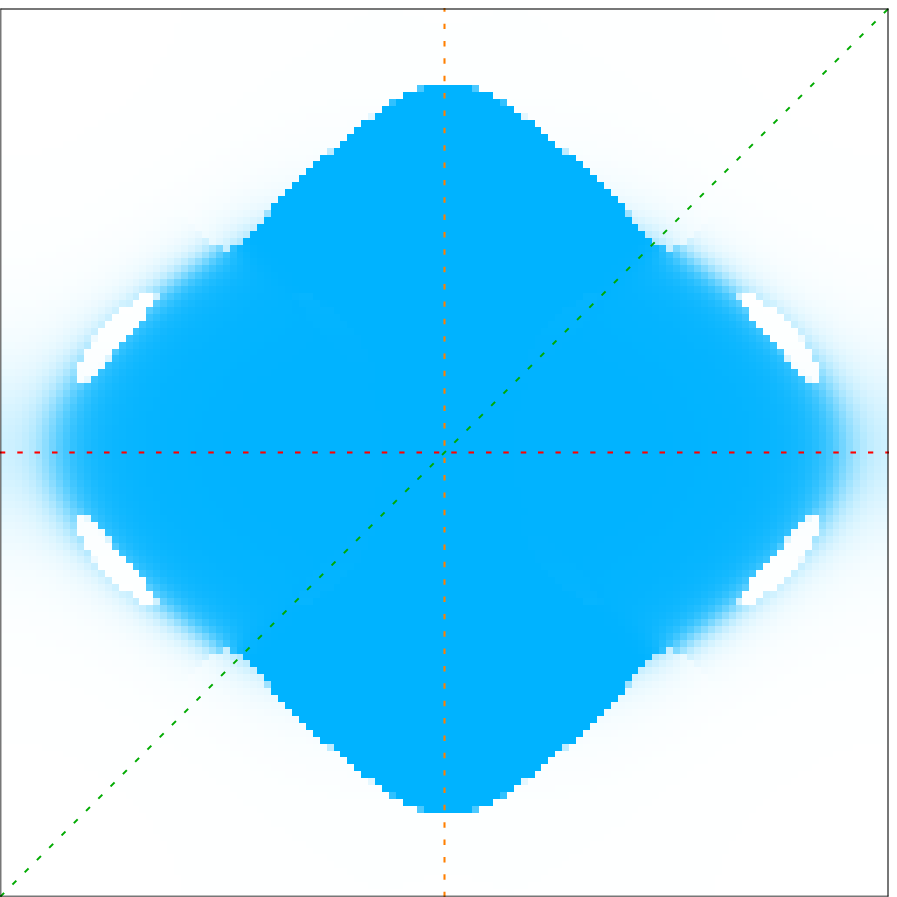}
}
\subfigure[$n_\dn(k_x,k_y,k_z=0)$]{
\includegraphics[width=0.43\columnwidth]{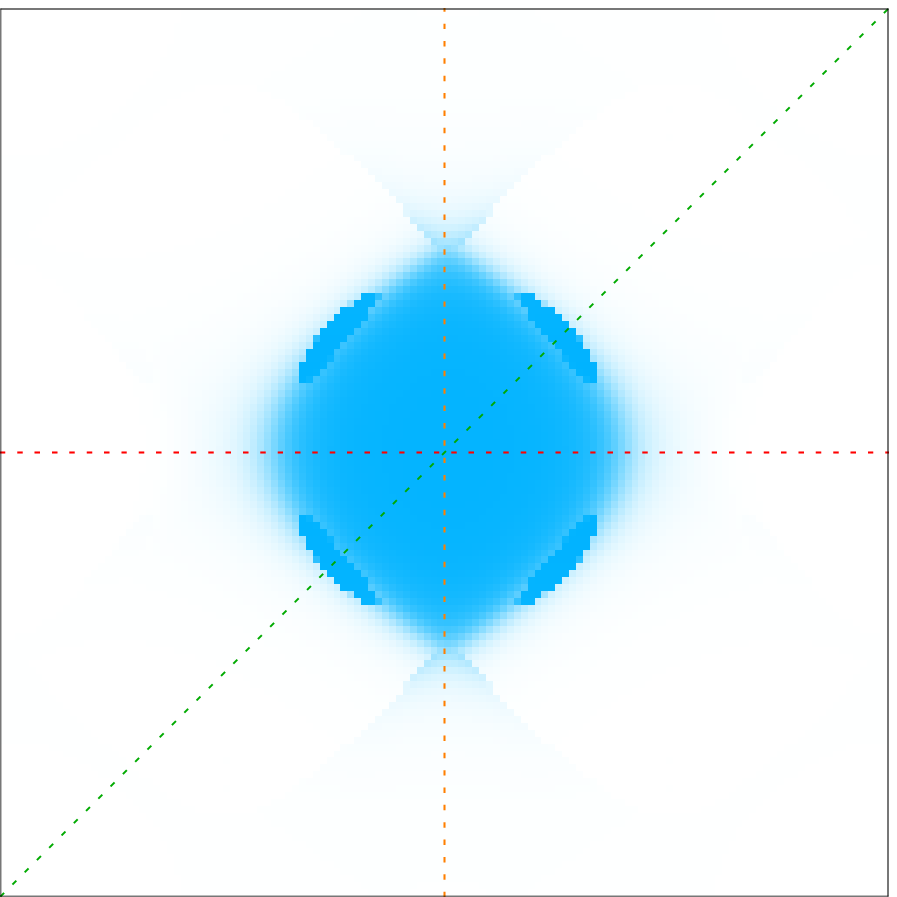}
}
\subfigure[$n_\up(\kkk)$ along slices]{
\includegraphics[width=0.45\columnwidth]{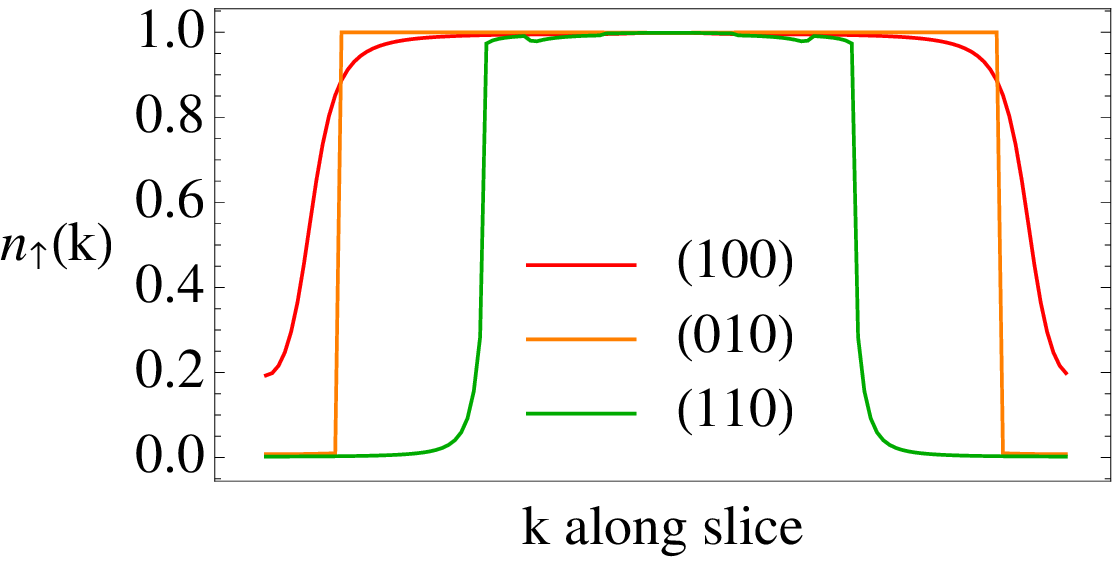}
}
\subfigure[$n_\dn(\kkk)$ along slices]{
\includegraphics[width=0.45\columnwidth]{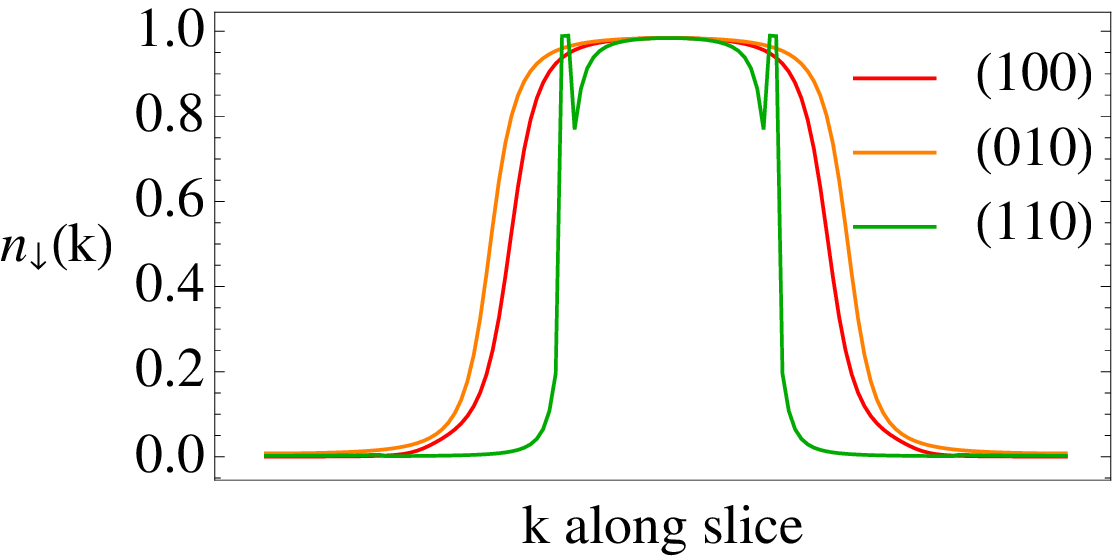}
}
\caption{
Momentum distributions of up- and down- fermions in an LO state with wavevector $\pm \qqq=(\pm\frac{2\pi}{4}, 0, 0)$. 	$n_\dn(\kkk)$ shows the effect of ``blocking'' \cite{fulde1964} by the sea of up spins $n_\up(\kkk)$ shifted by $\pm\qqq$, and vice versa.  The steps in the orange curve correspond to remnants of the up-spin Fermi surface.  The pattern breaks fourfold symmetry but preserves inversion symmetry, in contrast to Fig.~3 of Ref.~\onlinecite{koponen:120403}, where the $e^{i\qqq\cdot\rrr}$ ansatz breaks inversion symmetry.
\label{momentum-distributions}
}
\end{figure}
%%%%%%%%%% END FIGURE %%%%%%%%%%%%%%%%%

%==============================================================================
\paragraph{Nature of the LO state:}
%==============================================================================
% ($\Delta \sim \cos \qqq\cdot\rrr$) 
%In a lattice, anisotropy and commensurability effects introduce further complications.
%The $\mu=0$ line on the phase diagram is rather pathological because of a SU(2) degeneracy, so we will avoid this line, and stick to $\mu\geq 0.25$.  Then, we find that even if BdG is allowed to have a complex $\Delta$, it always converges to a state in which all $\Delta(\rrr)$ are real up to a global phase factor.

Figure~\ref{real-space} shows examples of weak and strong LO ground states in real space, and the textures in the repulsive Hubbard model that they correspond to under the LMT. 
At high fields just below $h_{LO}$, there is a weak LO state characterized by a sinusoidal order parameter with a $\qqq$-vector related to the difference between the Fermi wave vectors of the majority and minority components.  At lower fields this crosses over to a strong LO state consisting of BCS-like regions separated by domain walls\cite{burkhardt1994,yoshida:063601}.
The LO region actually includes a rich variety of patterns with different $\bf {q}$ (vertical stripes, diagonal stripes, 2D modulations) depending on commensurability effects.  

Our variational calculations find that FFLO ground states are always LO states that have a real order  parameter and break translational symmetry.  These states have a pairing energy that can be 50 times larger than the FF states studied in Ref.~\onlinecite{koponen:120403} that have a complex order parameter $\Delta \sim e^{i\qqq\cdot\rrr}$ and break time-reversal symmetry.  
This observation is consistent with previous results for the continuum.\cite{yoshida:063601,larkin1964}
According to the LMT, an LO state maps to a coplanar spin texture (spins in $xz$-plane), whereas an FF state maps to a non-coplanar ``helical'' texture.  Thus, for the \emph{repulsive} Hubbard model, coplanar textures are more favorable.

The pairing of up and down fermions belonging to unequal sized Fermi surfaces leads to complicated features in the momentum distribution function $n_\sigma(\kkk)$, such as Fermi arcs, Fermi pockets, and blocking regions.  Many patterns are possible depending on the parameters of the LO state.
Figure~\ref{momentum-distributions} shows an example of the momentum distributions, in which selective pairing produces non-monotonic behavior.  The most robust feature is the breaking of the lattice symmetry.  In experiments this effect may be complicated by twinning due to trap geometry.

%and difference $n_\up(\kkk)-n_\dn(\kkk)$, 

%%%%%%%%%% FIGURE (LaST FIG) %%%%%%%%%%%%%%%%%
\begin{figure*}
\subfigure[Phase diagram for $U=-6t$, $h=1.5t$]{
\includegraphics[width=0.30\textwidth]{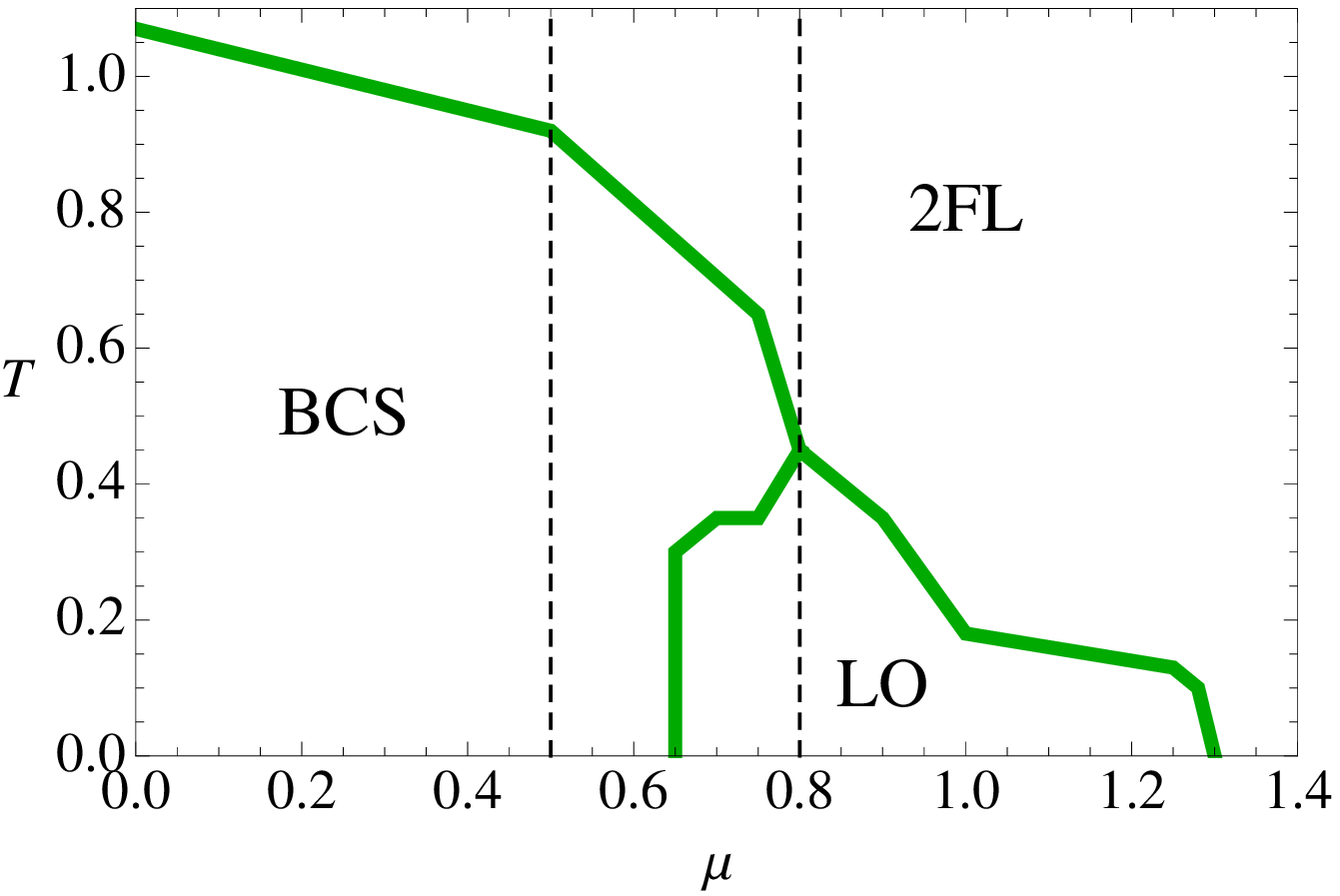}
\label{asdasd}
}
\subfigure[$\mu=0.5t$ (BCS g.s.)]{ %actually mu=-0.5, but doesn't matter
\includegraphics[width=0.30\textwidth]{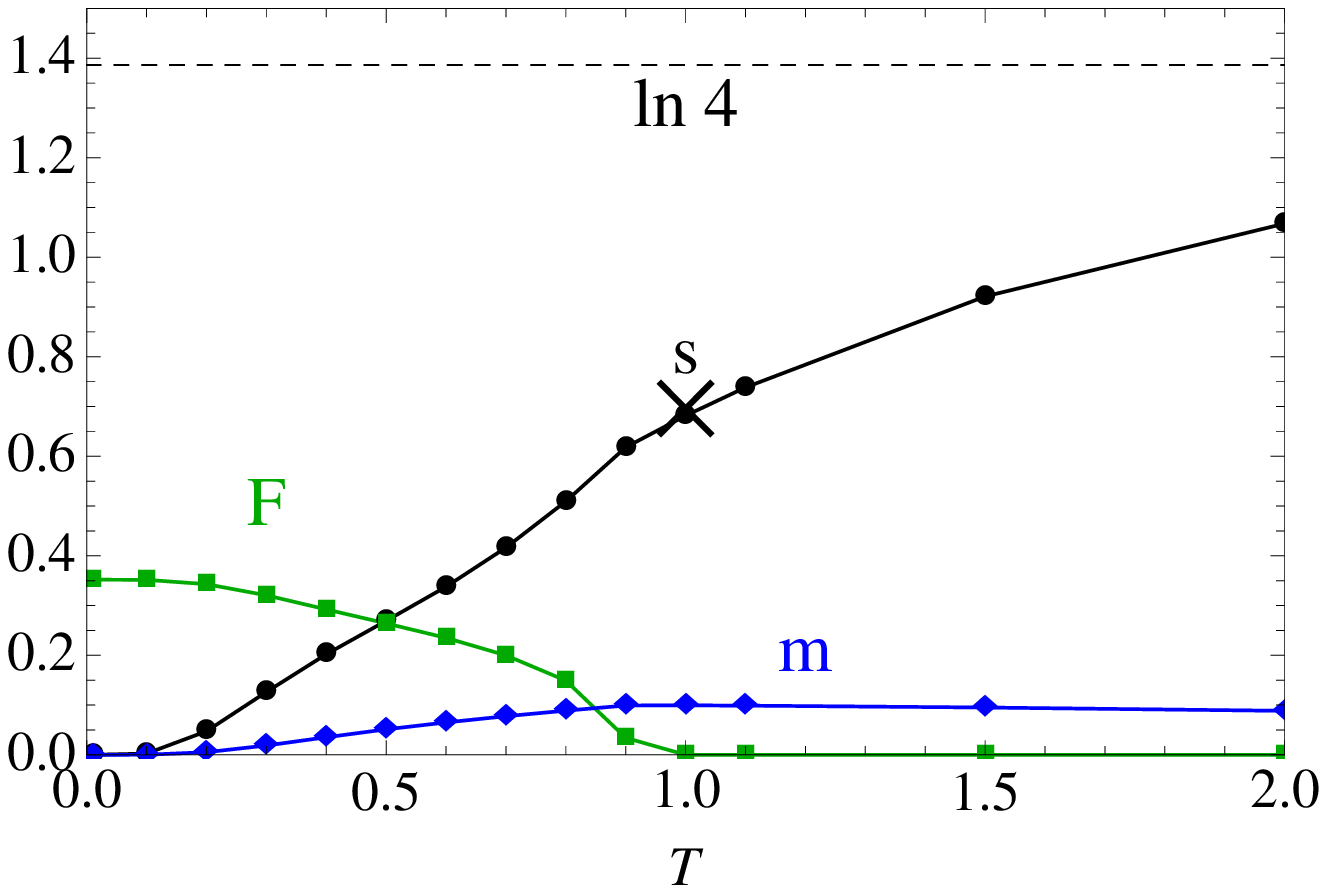}
}
\subfigure[$\mu=0.8t$ (LO g.s.)]{
\includegraphics[width=0.30\textwidth]{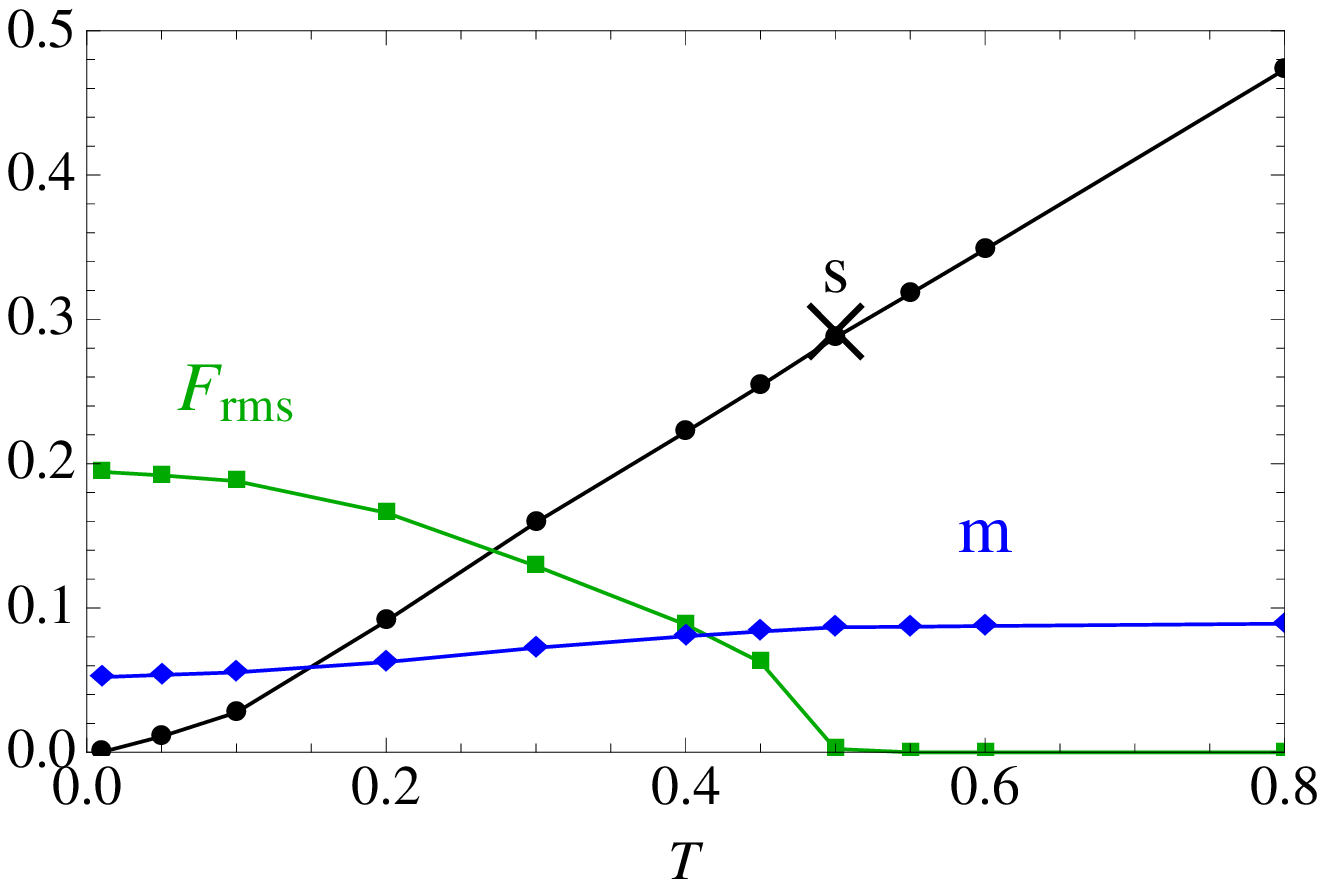}
}
\caption{
(a) A finite-temperature phase diagram.
(b) and (c): Entropy $s$ (per site), average polarization $m$, and pairing $F$ for $\mu=0.5t$ and $\mu=0.8t$ respectively, corresponding to dashed lines in (a).  
The BCS phase has $m=0$ at $T=0$; $m$ becomes non-zero at finite $T$ due to polarized quasiparticle excitations.  In contrast, the LO phase has $m>0$ even at $T=0$.
Note the different horizontal and vertical scales.
Crosses indicate critical temperatures and entropies ($s_{BCS}=0.69k_B$ and $s_{LO}=0.29k_B$); $s=k_B\ln 4=1.386k_B$ is the entropy of the Hubbard model at $T=\infty$.
\label{Tc}
}
\end{figure*}

%==============================================================================
\paragraph{Shell structure in a trap:}
%==============================================================================
%In the previous sections we have studied an infinite homogeneous system (or actually a box with periodic %boundary conditions). 
%In principle, densities $n_\sigma(\rrr)$ can be experimentally measured \emph{in situ} and  momentum %distributions $n_\sigma(\kkk)$ can be measured \emph{ex situ}.
%
%It is likely that the weak ``p-wave'' anisotropic component of the momentum distributions will be washed out by averaging over the entire gas.  As for real-space imaging of the density:  the density modulations in LO states are small, so high spatial resolution and high accuracy would be required.  
%
%We now combine our MFT with LDA to obtain predictions of the density profiles in a trap.
%For simplicity we consider harmonic spherical traps, although the same idea applies to cigar-shaped traps and other traps.

We now apply the above results for $n_\sigma(\rrr)$ and $n_\sigma(\kkk)$ to optical lattices in traps, within the local density approximation (LDA), which should be applicable to shallow traps with many fermions.
In LDA, the local phase is assumed to be determined by the local chemical potential, $\mu(r) = \mu_0 - V_\text{trap}(r)$.  This predicts shell structures corresponding to horizontal slices through the phase diagram (Fig.~\ref{U-6}); one such structure is depicted in Fig.~\ref{schematic}.
Spherical traps may cause twinning between LO states of different orientations, whereas a cigar-shaped trap helps align domain walls perpendicular to the long axis.  
In principle, 2FL--LO (and/or LO--BCS) transitions show up as kinks in the density profiles  $n_\up(\rrr)$ and $n_\dn(\rrr)$ with appropriate critical exponents; however, whether these kinks are observable depends on parameters and experimental resolution.

%  the functions $n$, $m$, and $f$ obey power laws with various critical exponents at each of the eight transitions, though these singularities will be broadened in finite traps.

%==============================================================================
\paragraph{Entropy for observing LO states:}
%==============================================================================
We predict that LO phases should be possible to observe at temperatures (or entropies) that are accessible to experiments.  
Within MFT we have found LO states up to $T_{LO}=0.6t$ ($s_{LO}=0.5k_B$), for the parameters 
$U=-6t, \mu=0.25t, h=2.25t$  This is not much lower than the critical temperature for the BCS phase, $T_{BCS} \approx 1.1t$ ($s_{BCS} \approx 0.8k_B$), at $U=-6t, \mu=0.25t, h=0$.
Ref.~\onlinecite{werner1995} implies that the critical entropy of the cubic lattice BCS phase in the $\left|U\right|\gg 12t$ regime is only reduced by about a factor of 2 from its mean-field value ($k_B\ln 2$).
Since our results are at medium coupling, $\left|U\right| = 6t$, we expect our MFT estimate of $s_{BCS}$ to be even closer to the true value.  
%Furthermore, since the dominant fluctuations in the BCS and LO phase involve $U(1)$ Goldstone modes, one may expect them to suppress the critical entropy by similar amounts.
Figure~\ref{Tc} compares the effect of temperature on BCS and LO phases.

In conclusion, we find that for the cubic lattice within fully self-consistent mean-field theory, LO states occur over an enhanced region of the phase diagram, as compared to the continuum.  
This suggests that imbalanced ultracold fermion systems in optical lattices should readily exhibit LO ground states, which could be detectable by virtue of the accompanying polarization oscillations.
Based on our calculations, we find that for $N \sim 10^5$ fermions with an overall polarization $P \sim 0.37$ at coupling $U=-6t$, about $83\percent$ of the atoms are in the LO phase.  
The polarization in each domain wall $P_{DW} = \sum_{\rrr\in DW} \frac{n_\up(\rrr) - n_\dn(\rrr)}{n_\up(\rrr) + n_\dn(\rrr)}$ for a strong LO state such as in Fig.~\ref{strong-LO-real-space.eps} is about $30\percent$; the polarization between domain walls is practically zero, giving a large contrast.  The spacing between domain walls can be of order $10a$, where $a$ is the optical lattice constant.  In optical lattice experiments with $a\sim 1\mathrm{\mu m}$, which implies a domain wall spacing of about $10 \mathrm{\mu m}$.

%(Further comments)
We explicitly describe the connection between LO phases in the attractive Hubbard model and spin textures in the repulsive Hubbard model via the LMT (a mapping that does not exist in the continuum).
Thus, by changing the sign of $U$, experiments in traps can effectively measure slices through the phase diagram in Fig.~\ref{U-6} in both horizontal and vertical directions.
We also point out that the strongly \emph{attractive} Hubbard model at half-filling and weak imbalance will have a tendency towards exotic $d$-wave magnetism described by order parameters such as $\mean{\cdag_{\kkk\up} \cccc_{\kkk\dn}} \sim \cos k_x - \cos k_y$, as a consequence of applying the LMT to the repulsive Hubbard model with its $d$-wave pairing tendencies.

Given that the enhanced LO regions in the cubic lattice occurs because of nesting, we expect that anisotropy will improve nesting and further enhance the LO region.  Other authors have studied 2D arrays of 1D tubes \cite{parish:250403,zhao:063605}; it remains to be shown whether anisotropic lattices or coupled tubes are more favorable for LO.  It will be important to include quantum  and thermal phase fluctuations in reduced dimensions to get accurate estimates of phase boundaries.

We acknowledge support from ARO and DARPA grant no. W911NF-08-1-0338.
\vspace{-0.5cm}

\bibliographystyle{forprl}
\bibliography{lo}

%%%%%%%%%%%%%%%%%%%%%%%%%%%%%%%%%%%%%%%%%%%%%%%
\end{document}